\documentclass{aastex62}

\received{...}
\revised{...}
\accepted{...}

\submitjournal{ApJ}

\shorttitle{On-disk reconnection condensations}
\shortauthors{Li et al.}

\begin{document}

\title{On-disk solar coronal condensations facilitated by magnetic reconnection between open and closed magnetic structures}

\correspondingauthor{Leping Li}
\email{lepingli@nao.cas.cn}

\author[0000-0001-5776-056X]{Leping Li}
\affil{CAS Key Laboratory of Solar Activity, National Astronomical Observatories, Chinese Academy of Sciences, Beijing 100101, People's Republic of China}
\affiliation{University of Chinese Academy of Sciences, Beijing 100049, People's Republic of China}

\author{Hardi Peter}
\affiliation{Max Planck Institute for Solar System Research, D-37077 G\"{o}ttingen, Germany}

\author[0000-0002-9270-6785]{Lakshmi Pradeep Chitta}
\affiliation{Max Planck Institute for Solar System Research, D-37077 G\"{o}ttingen, Germany}

\author[0000-0001-5705-661X]{Hongqiang Song}
\affiliation{Shandong Provincial Key Laboratory of Optical Astronomy and Solar-Terrestrial Environment, and Institute of Space Sciences, Shandong University, Weihai, Shandong 264209, People's Republic of China}

\begin{abstract}

Coronal condensation and rain are a crucial part of the mass cycle between the corona and chromosphere. In some cases, condensation and subsequent rain originate in the magnetic dips formed during magnetic reconnection. This provides a new and alternative formation mechanism for coronal rain. Until now, only off-limb, rather than on-disk, condensation events during reconnection have been reported. In this paper, employing extreme-ultraviolet (EUV) images of the Solar Terrestrial Relations Observatory (STEREO) and Solar Dynamics Observatory (SDO), we investigate the condensations facilitated by reconnection from 2011 July 14 to 15, when STEREO was in quadrature with respect to the Sun-Earth line. Above the limb, in STEREO/EUV Imager (EUVI) 171 \AA~images, higher-lying open structures move downward, reconnect with the lower-lying closed loops, and form dips. Two sets of newly reconnected structures then form. In the dips, bright condensations occur in EUVI 304 \AA~images repeatedly which then flow downward to the surface. In the on-disk observations by SDO/Atmospheric Imaging Assembly (AIA) in the 171 \AA~channel, these magnetic structures are difficult to identify. Dark condensations appear in AIA 304 \AA~images, and then move to the surface as on-disk coronal rain. The cooling and condensation of coronal plasma is revealed by the EUV light curves. If only the on-disk observations would be available, the relation between the condensations and reconnection, shown clearly by the off-limb observations, would not be identified. Thus, we suggest that some on-disk condensation events seen in transition region and chromospheric lines may be facilitated by reconnection.

\end{abstract}


\section{Introduction} \label{sec:int}

Condensation of hot solar coronal plasma is a widely observed phenomenon, which is best seen at the solar limb. Solar filaments/prominences are one of the prominent examples of coronal plasma condensation \citep{1983SoPh...88..219P, 2010SSRv..151..333M, 2014ApJ...792L..38X}. Formation of a prominence is generally initiated when coronal plasma is trapped in the dips of helical magnetic fields over polarity inversion lines. As the radiative losses from the trapped plasma exceed the heat input into the system, thermal instability is triggered, which further leads to rapid plasma cooling and condensation at coronal heights. The resulting condensed, dense plasma provides the mass of a prominence \citep{1953ApJ...117..431P, 1965ApJ...142..531F,1983SoPh...88..219P}. The extreme-ultraviolet (EUV) images recorded by the Atmospheric Imaging Assembly \citep[AIA;][]{2012SoPh..275...17L} onboard the Solar Dynamics Observatory \citep[SDO;][]{2012SoPh..275....3P}, revealed events of cooling and condensation of coronal plasma during the formation of prominence in a loop system \citep{2012ApJ...745L..21L} and a coronal cavity \citep{2012ApJ...758L..37B}, respectively. Similarly, coronal rain, which is seen as plasma draining through magnetic loops, is another manifestation of condensation \citep{2003A&A...411..605M, 2004A&A...424..289M, 2013ApJ...771L..29F, 2016ApJ...818..128O}. It appears frequently in post-flare loops and non-flaring active region (AR) closed loops \citep[see a review in][and references therein]{2020PPCF...62a4016A}. Quiescent coronal rain in AR closed loops forms due to thermal non-equilibrium when hot, dense plasma trapped by loops rapidly cools and condenses. Under the effect of gravity, the cool condensation falls down subsequently from the corona along one or both legs of the loops to the surface as coronal rain \citep{2003A&A...411..605M, 2004A&A...424..289M, 2015A&A...583A.109L, 2018ApJ...855...52F}. In this case, the hot, dense plasma could be supplied through heating events  concentrated at/near the loop footpoints \citep[][]{2017ApJ...835..272F,2018A&A...615L...9C}.  Alternatively, impulsive heating, occurring anywhere along a field line, could also trigger thermal instability and subsequent formation of coronal rain \citep{2019A&A...630A.123K}. Employing multi-wavelength images, the cooling and condensation events of coronal plasma in AR closed loops have been reported during the coronal rain formation \citep{2001SoPh..198..325S, 2012ApJ...745..152A, 2015A&A...577A.136V}.

One widely investigated concept of condensation relies on the thermal properties of the plasma alone, which is independent of the magnetic field, its topology and evolution \citep{2003A&A...411..605M, 2004A&A...424..289M}. In this scenario, the loss of thermal equilibrium between heat input, heat conduction, and radiative losses solely causes catastrophic cooling of plasma \citep{2013ApJ...771L..29F, 2014ApJ...792L..38X}. However, a reconnection-condensation model for prominence formation with in situ radiative condensation triggered by magnetic reconnection has been numerically simulated \citep{2015ApJ...806..115K, 2017ApJ...845...12K}. In line with these simulations , using SDO/AIA and Solar TErrestrail RElations Observatory \citep[STEREO;][]{2008SSRv..136....5K}/EUV Imager \citep[EUVI;][]{2008SSRv..136...67H} multi-wavelength images, we recently reported condensation events facilitated by reconnection between open and closed magnetic structures \citep{2018ApJ...864L...4L, 2018ApJ...868L..33L, 2019ApJ...884...34L, 2020ApJ...L}. Here, the magnetic reconnection is inferred from the reconfiguration of magnetic field topology as captured by the evolution of coronal structures, e.g., loops, as the magnetic flux is frozen into  plasma in the solar corona \citep{2000mrmt.conf.....P, 2009ApJ...703..877L, 2010ApJ...723L..28L, 2014ApJ...797L..14T, 2016Ap&SS.361..301L, 2016NatPh..12..847L, 2016ApJ...829L..33L, 2018ApJ...866...64C, 2018ApJ...853L..18Y, 2019ApJ...883..104S}. The simulations and observations suggest that the thermal evolution, e.g., coronal condensation, and the magnetic field evolution, e.g., magnetic reconnection, have to be treated together and cannot be separated \citep{2015ApJ...806..115K, 2017ApJ...845...12K, 2018ApJ...864L...4L, 2019ApJ...884...34L, 2020ApJ...L}. 

In a recent study, we presented observations that show a sequence of events from the formation of magnetic dips high in the corona to plasma condensations and subsequent rain along with signatures of reconnection \citep{2018ApJ...864L...4L}. This concerns with a system of magnetic structures, lasting for several days, out of the northwestern solar limb observed on 2012 January 19. The higher-lying open structures move down toward the surface, and reconnect with the lower-lying closed loops. Two sets of newly reconnected structures then form, and retract away from the reconnection region. During the reconnection process, a magnetic dip forms in the higher-lying open structures. Coronal plasma surrounding the dip converges into the dip, resulting in the enhancement of plasma density. In the dip, thermal instability of plasma, triggered by the density enhancement, occurs, and the coronal plasma rapidly cools and condenses. A transient prominence then forms, and facilitates a speed-up of the reconnection. Due to the successive reconnection, the condensation, without support from under-lying structures, falls down to the  surface along the newly reconnected closed loops, and also along the higher-lying open structures, as coronal rain. These observations form the basis for the proposal of a new and alternative mechanism for the formation of coronal rain along open field lines facilitated by interchange reconnection \citep{2018ApJ...864L...4L}. 

Formation of plasma condensations in the magnetic dips is coupled with quasi-periodic fast magnetoacoustic waves that originate from the reconnection region and propagate upward across the dips of higher-lying open structures \citep{2018ApJ...868L..33L}. This indicates that most of the magnetic energy may be converted to the wave energy through reconnection. Such repeated condensation events facilitated by reconnection between open and closed structures are observed over an extended period of time \citep[e.g., 15 events in the long-term evolution of the same magnetic structure system during 2012 January 16 to 20; see][]{2019ApJ...884...34L}. However, no condensation, and thus no subsequent coronal rain, is detected in one of the repeated reconnection events. This suggests that not each reconnection event will lead to the formation of condensation \citep{2019ApJ...884...34L}. Moreover, 79 other similar reconnection and condensation events are identified above the limb at different times and different places in the month of 2012 January, employing the AIA observations with a much longer time cadence of 14 minutes \citep{2019ApJ...884...34L}. The condensation, and hence the coronal rain, facilitated by reconnection is a common, rather than a special, phenomenon in the corona. Recently, using Interface Region Imaging Spectrograph \citep[IRIS;][]{2014SoPh..289.2733D} and AIA images and spectra, we analyze a coronal rain event with loop-like paths in the transition region and chromospheric lines, e.g., Si IV 1394 \AA~and Mg II k 2796 \AA, and find that it originates from the condensations facilitated by reconnection between open and closed structures \citep{2020ApJ...L}. Some of the coronal rain events observed at chromospheric temperatures thus could be explained by the formation mechanism of coronal rain, where the condensation is facilitated by interchange reconnection. Moreover, the occurrence interval of repeated coronal rain events is obtained to be in the range of 4.6-10.4 hr with a mean value of 6.6 hr \citep{2020ApJ...L}. This recurrence interval is identical with the periods of quiescent coronal rain in AR closed loops \citep{2018ApJ...853..176A, 2020A&A...633A..11F}. However, different from the periodicity of coronal rain in AR closed loops that is driven by thermal non-equilibrium, the global topological changes caused by reconnection clearly play a role in this repetition, as the repetition observed in the coronal rain along open structures closely follows the formation of dips \citep{2020ApJ...L}.

Coronal rain events occurring at/near null points and associated dips have been reported \citep{2012ApJ...745L..21L, 2015ApJ...807....7R, 2018ApJ...864L...4L, 2018ApJ...868L..33L, 2019ApJ...884...34L, 2020ApJ...L, 2019ApJ...874L..33M}. They belong to quiescent coronal rain, as no associated flare is observed during the reconnection and condensation process \citep{2018ApJ...864L...4L, 2018ApJ...868L..33L, 2019ApJ...884...34L, 2020ApJ...L}. Quite different from the quiescent coronal rain events along closed loops in non-flaring AR reported previously \citep{2001SoPh..198..325S, 2012ApJ...745..152A, 2015A&A...577A.136V, 2019A&A...630A.123K}, the quiescent coronal rain here occurs in the dips of open structures. Because of this, their dynamics at coronal heights are quite different \citep[see more details in][]{2020ApJ...L}. The main difference is that in the case of coronal rain occurring in AR closed loops, it is mostly observed to fall down toward the surface along one or both loop legs \citep{2020PPCF...62a4016A}. In contrast, condensation that forms in the dips of open structures accumulates, expands, and flows along the supporting field as a prominence, and moves together with the evolution of dips, before it falls down toward the surface \citep{2018ApJ...864L...4L, 2019ApJ...884...34L, 2020ApJ...L}. Furthermore, the coronal rain in AR closed loops is driven by thermal non-equilibrium promoted by footpoint-concentrated heating events \citep{2003A&A...411..605M, 2004A&A...424..289M, 2020PPCF...62a4016A}. On open field lines, \citet{2019ApJ...874L..33M} suggested that thermal non-equilibrium is not expected to occur. In contrast, in their 3D magnetohydrodynamic (MHD) models \citet{2020A&A...639A..20K} found condensations also on magnetic field lines that reach the top of the computational domain at 14.4 Mm above the photosphere and can therefore be considered locally open. Here the loss of thermal equilibrium seems to be induced in a similar fashion as in closed loops, i.e., by the gradual increase of density due to an evaporated upflow driven by footpoint heating. This magnetic configuration of \citet{2020A&A...639A..20K} is quite different from large-scale open structures with dips where a local density enhancement through reconnection facilitates the loss of thermal equilibrium \citep{2018ApJ...864L...4L, 2018ApJ...868L..33L, 2019ApJ...884...34L, 2020ApJ...L, 2019ApJ...874L..33M}. All these results point to a new class of quiescent coronal rain along open structures in comparison with that in AR closed loops \citep[see more details in][]{2020ApJ...L}. 

Until now, owing to the ease of detection, the reported condensation events facilitated by reconnection between open and closed structures are all seen above the solar limb \citep{2018ApJ...864L...4L, 2018ApJ...868L..33L, 2019ApJ...884...34L, 2020ApJ...L, 2019A&A...630A.123K, 2019ApJ...874L..33M}. Whether the condensation facilitated by reconnection can still be observed further on the disk, and how it performs are open questions. To answer these questions, in this study, we investigate the condensation events facilitated by reconnection both above the limb and on the disk. Comparing with the off-limb events, how the magnetic fields, associated with the magnetic structure system, distribute and evolve can be investigated by employing the on-disk events. The observations and method are described in Section \ref{sec:obs}, the results are shown in Section \ref{sec:res}, and a summary and discussion are given in Section \ref{sec:sum}. 

\section{Observations and method}\label{sec:obs}

AIA onboard the SDO is a set of normal incidence imaging telescopes designed to acquire images of the solar atmosphere at ten wavelength bands. EUVI onboard the STEREO provides solar images in four EUV channels of 171 \AA, 195 \AA, 284 \AA, and 304 \AA. Here, different channels show plasma at different temperatures, e.g., 171 \AA~peaks at $\sim$0.9 MK (Fe IX), 131 \AA~peaks at $\sim$0.6 MK (Fe VIII) and $\sim$10 MK (Fe XXI), and 304 \AA~peaks at $\sim$0.05 MK (He II). In this study, we employ the AIA 171 \AA, 131 \AA, and 304 \AA~images, and the EUVI A and B 171 \AA~and 304 \AA~images to investigate the evolution of magnetic structures and condensations. Here, to better show the evolution, the EUVI A and B 171 \AA~images and AIA 304 \AA~images are enhanced by using the Multi-scale Gaussian Normalization (MGN) technique \citep{2014SoPh..289.2945M}. Spatial sampling of the employed AIA and EUVI images is 0.6\arcsec/pixel and 1.6\arcsec/pixel, respectively. Time cadence of the AIA EUV images is 12 s, while that of the EUVI images is non-uniform. Helioseismic and Magnetic Imager \citep[HMI;][]{2012SoPh..275..229S} line of sight (LOS) magnetograms onboard the SDO, with time cadence and spatial sampling of 45 s and 0.5\arcsec/pixel, are also used to study the evolution of associated photospheric magnetic fields.

\begin{figure}[ht!]
\centering
\includegraphics[width=0.88\textwidth]{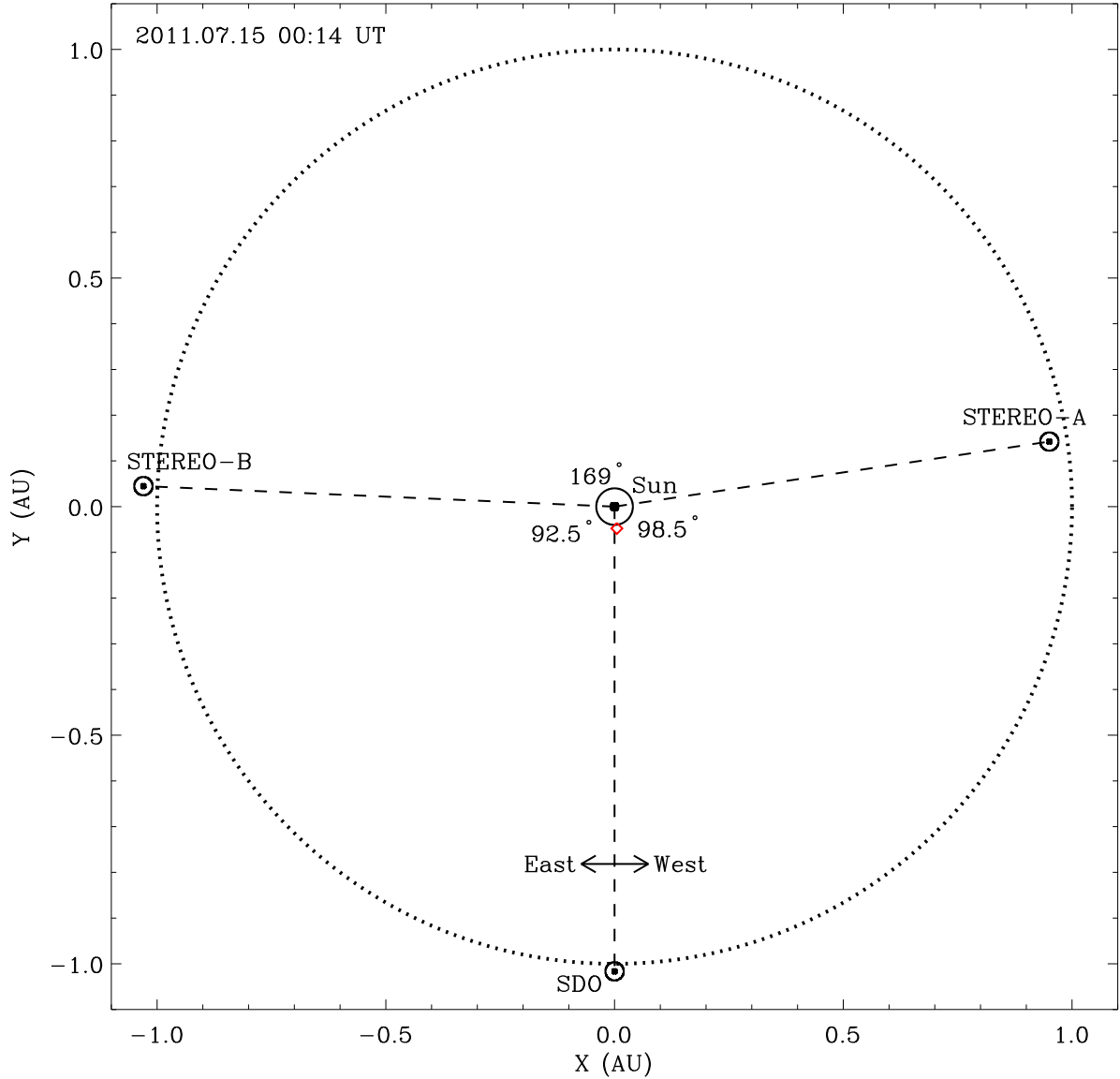}
\caption{Positions of the STEREO A and B and SDO satellites at 00:14 UT on 2011 July 15. The angles between these three satellites are denoted by the numbers in the plot. The dotted circle represents the Earth orbit at 1 AU. East and West separately show the east and west directions in the field of view (FOV) of SDO. The red diamond marks the general location of coronal structures we studied. See Section \ref{sec:obs} for details. 
\label{f:satellites}}
\end{figure}

The SDO and STEREO A and B satellites observe the Sun generally at different angles. The positions of these three satellites at 00:14 UT on 2011 July 15 are shown in Figure \ref{f:satellites}. In order to find the on-disk condensations facilitated by reconnection between open and closed structures in the FOV of SDO, we choose a period of time, e.g., from 2010 September to 2011 September, when the viewing direction of SDO is mostly perpendicular to those of STEREO A and B. During this time period, we examine the evolution of magnetic structures out of the eastern (western) limb by using EUVI A (B) 171 \AA~and 304 \AA~images, to find the off-limb condensation events facilitated by reconnection in the FOV of STEREO A (B). Under an almost orthogonal viewing angle of SDO, the off-limb condensation events facilitated by reconnection in STEREO A and B FOVs are hence located on the disk in the FOV of SDO. In this study, the condensation events facilitated by reconnection between open and closed structures from 2011 July 14 to 15 above the southeastern (southwestern) limb in EUVI A (B) images are chosen for analysis (see the red diamond in Figure \ref{f:satellites}). In this time period, time cadence of EUVI A (B) 171 \AA~images ranges from 30 s to 258 (120) minutes with the most value of 75 s, and that of EUVI A (B) 304 \AA~images varies from 2.5 minutes to 250 (10) minutes with two most values of 2.5 minutes and 10 minutes.

\section{Results}\label{sec:res}

\subsection{Off-limb coronal condensations facilitated by reconnection between open and closed structures observed by STEREO}\label{sec:cc_stereo}

From 2011 July 14 to 15, the long-term evolution of a set of higher-lying open structures L1 is observed above the southwestern (southeastern) limb in EUVI B (A) 171 \AA~images, see Figures \ref{f:general}(a)-(b). The NOAA AR 11250 is located to the north of the structures, see Figures \ref{f:general}(a)-(b). A filament lies underneath the structures L1, see Figure \ref{f:general}(b). The structures L1 move down toward the surface, with obvious structural changes, and interact with the lower-lying closed loops, L2, see Figures \ref{f:cc_stereo}(a) and (d) and the online animated version of Figure \ref{f:cc_stereo}. Along the AB direction in the red rectangle in Figure \ref{f:cc_stereo}(a), a time slice of EUVI B 171 \AA~images is made, and displayed in Figure \ref{f:measurements_stereo}(a). The downward motion of structures L1 is clearly detected perpendicular to the line of sight with a mean speed of 0.7 km s$^{-1}$, see the red dotted line in Figure \ref{f:measurements_stereo}(a). The structures L1 and L2 interact and give rise to a pair of new structures L3 and L4. This process is consistent with magnetic reconnection at the interface of structures L1 and L2, see Figures \ref{f:cc_stereo}(a) and (d) and the online animated version of Figure \ref{f:cc_stereo}.

Due to the downward motion of structures L1, a magnetic dip forms in structures L1, see Figures \ref{f:cc_stereo}(a) and (d). Condensations of coronal plasma appear in EUVI A and B 304 \AA~images at the north edge of the dip. They then flow down along both legs of the newly reconnected closed loops L4, and also the leg of higher-lying open structures L1, to the surface as coronal rain, and finally disappear, see Figures \ref{f:cc_stereo}(b) and (e) and the online animated version of Figure \ref{f:cc_stereo}. Along the CD direction in the green box in Figure \ref{f:cc_stereo}(b), a time slice of EUVI B 304 \AA~images is made and shown in Figure \ref{f:measurements_stereo}(b). Multiple flows of the condensations are evidently identified with speeds ranging from 15 km s$^{-1}$ to 30 km s$^{-1}$, see the green dotted lines in Figure \ref{f:measurements_stereo}(b). 

During the two days, condensations take place repeatedly for five times. General information on the repeated condensation events is listed in Table \ref{tab:information}. Parameters, including the start time, the end time, the appearance time, the disappearance time, and the lifetime of the condensations are measured and listed in Table \ref{tab:information}. The lifetimes of condensations range from 1.5 hr to 10.2 hr. Because the time cadences of EUVI A and B 304 \AA~images here are non-uniform, those surrounding the measured parameters are chosen as measurement errors, e.g., $\pm$0.2 hr is taken as the measurement error of the appearance time in EUVI A 304 \AA~images for the first condensation event, see Table \ref{tab:information}. Additionally, due to the longer, non-uniform time cadences of EUVI A and B 304 \AA~images, downward flow of the second condensation event along the north leg of loops L4 and/or the leg of structures L1 looks like a chromospheric surge, see the online animated version of Figure \ref{f:cc_stereo}. We carefully check the simultaneous  associated AIA 304 \AA~images, see the online animated version of Figure \ref{f:cc_sdo}, and only detect the northward, rather than the southward, flows, see Section \ref{sec:cc_sdo}. This indicates that the coronal rain, rather than the chromospheric surge, takes place during the second condensation event. In the intervals between any two neighboring reconnection and condensation events, the higher-lying open structures L1 remain quiescent, and show inclined structures, see Figures \ref{f:cc_stereo}(c) and (f). Without reconnection between the open and closed structures, no formation of dip in the higher-lying open structures, and no cooling and condensation of coronal plasma is observed \citep{2018ApJ...864L...4L, 2018ApJ...868L..33L, 2019ApJ...884...34L, 2020ApJ...L}.

In the purple rectangles in Figures \ref{f:cc_stereo}(a)-(b), the light curves of the EUVI B 171 \AA~and 304 \AA~channels for the second reconnection and condensation event are calculated and displayed in Figure \ref{f:measurements_stereo}(c) as blue and purple squares and lines. Here, we show the observed values as squares on the lines, clearly showing the time cadences of EUVI B images. In the blue rectangle in Figure \ref{f:cc_stereo}(d), the light curve of the EUVI A 171 \AA~channel is measured and shown in Figure \ref{f:measurements_stereo}(c) as green diamonds and line. The light curve of the EUVI A 304 \AA~channel is, however, not calculated, as the time cadence of EUVI A 304 \AA~images during the time period is non-uniform. The EUVI B and A 171 \AA~light curves increase, reach the peaks, and then decrease, see the blue and green lines in Figure \ref{f:measurements_stereo}(c). They peak at 18:14 UT on 2011 July 14, see the vertical blue dotted and green dashed lines in Figure \ref{f:measurements_stereo}(c). Similar to the EUVI B and A 171 \AA~light curves, the light curve of the EUVI B 304 \AA~channel also increases, reaches the peak, and then decreases. It, however, peaks at 20:36 UT on 2011 July 14, about 2.4 hr after the peaks of the EUVI B and A 171 \AA~light curves, see the purple vertical dotted line in Figure \ref{f:measurements_stereo}(c). This progressive appearance of emission first in the 171\,\AA\ channel and then in the 304\,\AA\ channel is consistent with plasma cooling and condensation \citep{2012ApJ...753...35V}. The plasma in the dip of structures L1 cools down from $\sim$0.9 MK, the characteristic temperature of the 171 \AA~channel, to $\sim$0.05 MK, the characteristic temperature of the 304 \AA~channel, in about 2.4 hr for the EUVI B. The peak times of the EUVI A 171 \AA, B 171 \AA, and B 304 \AA~light curves, and the cooling times from 171 \AA~channel to 304 \AA~channel are listed in Table \ref{tab:measurements}. Similarly, the time cadences surrounding the measured parameters are considered as measurement errors, see Figure \ref{f:measurements_stereo}(c) and  Table \ref{tab:measurements}. Taking the measurement errors into account, identical results of the parameters are obtained for both the EUVI A and B, see Table \ref{tab:measurements}. For the other reconnection and condensation events, similar evolution of the EUVI 171 \AA~and 304 \AA~light curves is obtained. In this study, even though the time cadences of EUVI A and B observations are longer and non-uniform, the signatures of reconnection and condensations are clearly identified, see Figures \ref{f:cc_stereo} and \ref{f:measurements_stereo} and the online animated version of Figure \ref{f:cc_stereo}.

\subsection{On-disk coronal condensations facilitated by reconnection between open and closed structures observed by SDO}\label{sec:cc_sdo}

In consideration of the separation angles between the three satellites of SDO and STEREO A and B, see Figure \ref{f:satellites}, combining the evolution of the 171 \AA~structures in EUVI A and B, we identify the same structures in AIA 171 \AA~images on the disk, see Figure \ref{f:general}(c). The identification is confirmed as the observed structures in the AIA 171\,\AA\ channel are located such that the AR 11250 is located to the north of the structures L1, and the filament is lying underneath the structures L1, see Figure \ref{f:general}(c). As the structures are located on the disk in the FOV of SDO, their associated photospheric magnetic fields are thus observed in HMI LOS magnetograms, see Figure \ref{f:general}(d). The open structures L1 and L3 root in positive magnetic field concentrations, enclosed separately by the blue circles and ellipses in Figure \ref{f:cc_sdo}. For the magnetic fields enclosed by the blue circles, many dynamics, e.g., the motion, separation, convergence, and rotation, are detected, see the online animated version of Figure \ref{f:cc_sdo}. Moreover, a mean magnetic field strength of 100 G and a total magnetic flux of 4$\times$10$^{11}$ Mx are measured for the structures L1. The filament, underneath the structures L1, is located above the polarity inversion line between opposite polarity magnetic fields, see the cyan dotted lines in Figures \ref{f:cc_sdo}(e) and (g). The loops L2 are difficult to observe on the disk in AIA 171 \AA~images. According to the EUVI A and B 171 \AA~observations, see Section \ref{sec:cc_stereo}, they should connect the positive magnetic field concentration, the endpoint of structures L3, and the negative magnetic field concentrations, located between the endpoints of open structures L1 and L3. Moreover, the endpoint of loops L2, rooting in the negative magnetic field concentrations, is located closer to the other endpoint of loops L2 than to the endpoint of structures L1.

Due to the SDO viewing angle, the almost line-of-sight downward motion of structures L1 is difficult to detect in AIA images. Nevertheless, a bright emission feature appears at the north edge of the dip of structures L1 in AIA 171 \AA~and 131 \AA~images, marked by blue solid arrows in Figures \ref{f:cc_sdo}(a) and (c). Here, the AIA 131 \AA~emission shows plasma with the lower characteristic temperature ($\sim$0.6 MK) of the AIA 131 \AA~channel, because there is no associated bright emission identified in the AIA higher-temperature channels, e.g., 94 \AA, 335 \AA, 211 \AA, and 193 \AA. Subsequently, the leg of structures L1 appears, see Figures \ref{f:cc_sdo}(a) and (c) and the online animated version of Figure \ref{f:cc_sdo}. We measure the width of structures L1 in the dip region, and get a value of $\sim$22 Mm. Assuming that the structures L1 are cylinder in the dip region, with a diameter of 22 Mm, we calculate the magnetic field strength of structures L1 in the dip region using the total magnetic flux of structures L1 (4$\times$10$^{11}$ Mx), and obtain a mean value of 10 G. The newly reconnected open structures L3 then form, see Figures \ref{f:cc_sdo}(b) and (d). At the same time, part of the leg of structures L1 turns into the north leg of the newly reconnected closed loops L4 through reconnection with loops L2. However, they are not distinguished from each other, due to the viewing angle of SDO, see Figures \ref{f:cc_sdo}(b) and (d). 

Dark condensations are detected in AIA 304 \AA~images, with the width of $\sim$1.4\arcsec, that flow along the leg of structures L1 and the north leg of loops L4 (toward their north endpoints), denoted by green solid arrows in Figures \ref{f:cc_sdo}(e)-(f). Similar as in the STEREO observations, cooling and condensation of coronal plasma in the dip is thus identified on the disk with AIA. In order to better show the condensations, the AIA 304 \AA~image in the cyan rectangle in Figure \ref{f:cc_sdo}(e) is enlarged in Figure \ref{f:measurements_sdo}(a), see the online animated version of Figure \ref{f:measurements_sdo}. Here, the blue dash-dotted lines in Figures \ref{f:measurements_sdo}(a)-(b) enclose the AIA 131 \AA~structures L1 and L4, showing the moving path of condensations. Along the green dotted line EF in Figure \ref{f:measurements_sdo}(a), a time slice of AIA 304 \AA~images is made and displayed in Figure \ref{f:measurements_sdo}(c). Multiple flows of the condensations are identified, denoted by green solid arrows in Figure \ref{f:measurements_sdo}(c), with moving speeds ranging from 37 km s$^{-1}$ to 72 km s$^{-1}$, see the cyan dashed lines in Figure \ref{f:measurements_sdo}(c). Here, besides the dark features, some bright features are also detected, see Figure \ref{f:measurements_sdo}(c). This may be caused by the multithermal character of coronal rain. 

In the blue rectangles in Figures \ref{f:cc_sdo}(b), (d), and (f), the light curves of the AIA 171 \AA, 131 \AA, and 304 \AA~channels are calculated and displayed in Figure \ref{f:lightcurves_sdo} as red, green, and blue lines. Here, the inverse of the AIA 304 \AA~light curve is used, as the condensations in AIA 304 \AA~images show dark absorption features. All the three AIA EUV light curves increase, reach the peaks, and then decrease. They, however, peak separately at 19:35 UT, 20:08 UT, and 20:34 UT on 2011 July 14, marked by the red, green, and blue vertical dotted lines in Figure \ref{f:lightcurves_sdo}. The fact that the AIA 131 \AA~light curve peaks later than the AIA 171 \AA~light curve also supports that the AIA 131 \AA~emission represents plasma with the lower characteristic temperature ($\sim$0.6 MK) of the AIA 131 \AA~channel, cooler than the emitting plasma detected by the AIA 171 \AA~channel ($\sim$0.9 MK). The cooling of plasma, always shown by the AIA EUV light curves in the corona \citep{2012ApJ...753...35V}, in the dip of open structures L1 is hence clearly detected. Here, the plasma cools down from $\sim$0.9 MK, the characteristic temperature of the AIA 171 \AA~channel, to $\sim$0.6 MK, the lower characteristic temperature of the AIA 131 \AA~channel, in 33 minutes, and then to $\sim$0.05 MK, the characteristic temperature of the AIA 304 \AA~channel, in another 26 minutes, see the second condensation event in Table \ref{tab:measurements}. Similar evolution of the AIA EUV light curves is obtained for the other reconnection and condensation events. For the five condensation events, the plasma in the dip cools down from $\sim$0.9 MK to $\sim$0.6 MK in 0.13-1.05 hr, and then to $\sim$0.05 MK in another 0.43-2.75 hr, see Table \ref{tab:measurements}. More details on the repeated reconnection and condensation events above the limb are revealed in \citet{2019ApJ...884...34L, 2020ApJ...L}. To compare with the STEREO observations, we overlay the EUVI B 304 \AA~light curve in Figure \ref{f:measurements_stereo}(c) on Figure \ref{f:lightcurves_sdo} as purple squares and line. Even though the condensation in AIA 304 \AA~images shows dark absorption feature, and is affected by the underneath background emission, the AIA 304 \AA~light curve shows similar temporal evolution to the light curve of the EUVI B 304 \AA~channel, see Figure \ref{f:lightcurves_sdo}. Both of them peak at almost the same time. In consideration of the measurement errors, consistent parameters, i.e., the peak and cooling times, are obtained from the AIA and EUVI A and B observations, see Table \ref{tab:measurements}.

\section{Summary and discussion}\label{sec:sum}

Employing EUV images and LOS magnetograms of the STEREO A and B and SDO, we investigate the condensation events facilitated by reconnection between open and closed structures both above the limb and on the disk from 2011 July 14 to 15. In EUVI B (A) 171 \AA~images, the higher-lying open structures L1 above the southwestern (southeastern) limb move down toward the surface, with the formation of a magnetic dip, and interact with the lower-lying closed loops L2. Reconnection between the structures L1 and L2 takes place at the interface, and forms the newly reconnected structures L3 and L4. Five bright condensation events appear repeatedly in EUVI A and B 304 \AA~images at the north edge of the dip of structures L1, and move downward to the surface along both legs of loops L4 and the leg of structures L1. In AIA 171 \AA~images, due to the viewing angle, the whole system of reconnection structures is not totally detected on the disk. Only part of the structures L1, L3, and L4 are identified. Bright emission appears in AIA 171 \AA~and 131 \AA~images at the north edge of the dip of structures L1. Dark condensations subsequently occur in AIA 304 \AA~images, and then flow downward to the surface along the north leg of structures L1 and L4. The cooling and condensation process of coronal plasma in the dip of structures L1 is evidently observed, also identified from the EUVI 171 \AA~and 304 \AA~and the AIA 171 \AA, 131 \AA, and 304 \AA~light curves. 

According to the EUVI A and B and AIA EUV images and the HMI LOS magnetograms, schematic diagrams, adapted from Figure 6 of \citet{2018ApJ...864L...4L}, are provided to describe the reconnection between structures and the condensations of plasma from three different observing angles in Figure \ref{f:cartoon}. Reconnection between the field lines of structures L1 and L2, see red stars between the green and blue lines in Figure \ref{f:cartoon}, forms a magnetic dip in structures L1, marked by red solid arrows in Figure \ref{f:cartoon}. The surrounding plasma is gathered into the dip, leading to the enhancement of plasma density. Triggered by the density enhancement, thermal instability takes place locally in the dip, driving the plasma to cool catastrophically and condense, see large orange ellipses in Figure \ref{f:cartoon}. Due to the successive reconnection, the condensations fall down along the field lines of structures L1 and L4 to the surface as coronal rain, see small orange ellipses along the lines in Figure \ref{f:cartoon}.

\begin{figure}[ht!]
\centering
\includegraphics[width=0.88\textwidth]{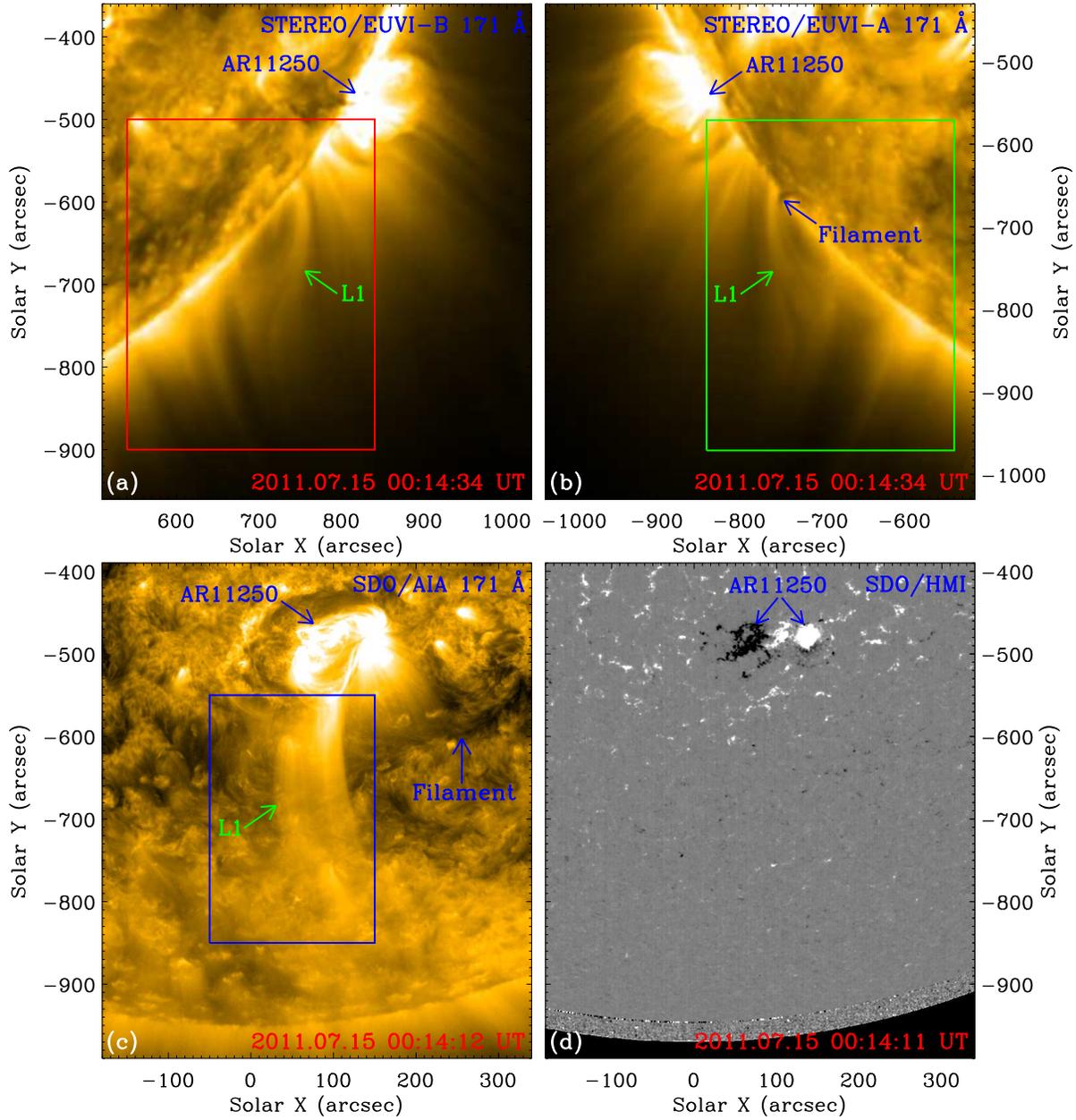}
\caption{Overview of the structures observed by STEREO A and B and SDO. (a) STEREO/EUVI B and (b) A and (c) SDO/AIA 171 \AA~images, and (d) a SDO/HMI LOS magnetogram. The red, green, and blue rectangles in (a), (b), and (c) denote the FOVs of Figures \ref{f:cc_stereo}(a)-(c), \ref{f:cc_stereo}(d)-(f), and \ref{f:cc_sdo}, respectively. See Sections \ref{sec:cc_stereo} and  \ref{sec:cc_sdo} for details.
\label{f:general}}
\end{figure}

\begin{figure}[ht!]
\includegraphics[width=0.88\textwidth]{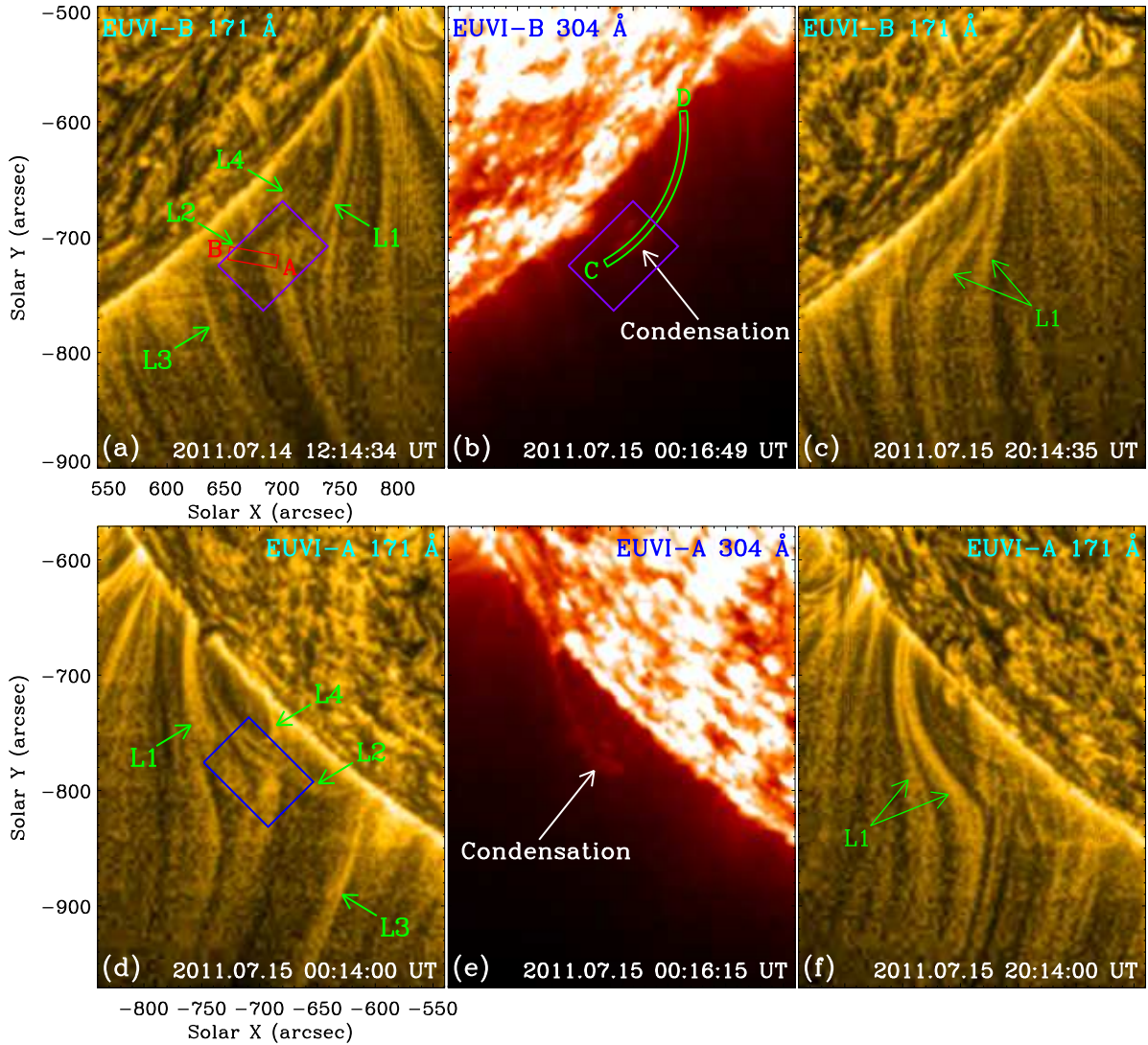}
\centering
\caption{Coronal condensations facilitated by magnetic reconnection between open and closed structures observed by STEREO A and B. (a)-(c) and (d)-(f) EUVI B and A 171 \AA~and 304 \AA~images. Here, the EUVI B and A 171 \AA~images in (a), (c)-(d), and (f) have been enhanced using the MGN technique. The red rectangle AB in (a) and the green box CD in (b) mark the positions of features used to construct time slices of EUVI B 171 \AA~and 304 \AA\ diagnostics displayed in Figures \ref{f:measurements_stereo}(a) and (b). The purple rectangles in (a)-(b) mark the regions for the light curves of the EUVI B 171 \AA~and 304 \AA~channels as shown in Figure \ref{f:measurements_stereo}(c) by the blue and purple squares and lines. The blue rectangle in (d) enclose the region for the EUVI A 171 \AA~light curve as displayed by the green diamonds and line in Figure \ref{f:measurements_stereo}(c). The FOVs of (a)-(c) and (d)-(f) are denoted separately by the red and green rectangles in Figures \ref{f:general}(a) and (b). An animation of the unannotated EUVI images is available. It covers $\sim$2 days starting at 00:07 UT on 2011 July 14, and the time cadence varies from 30 s to 4.3 hr. See Section \ref{sec:cc_stereo} for details. (An animation of this figure is available.)
\label{f:cc_stereo}}
\end{figure}

\begin{figure}[ht!]
\includegraphics[width=0.9\textwidth]{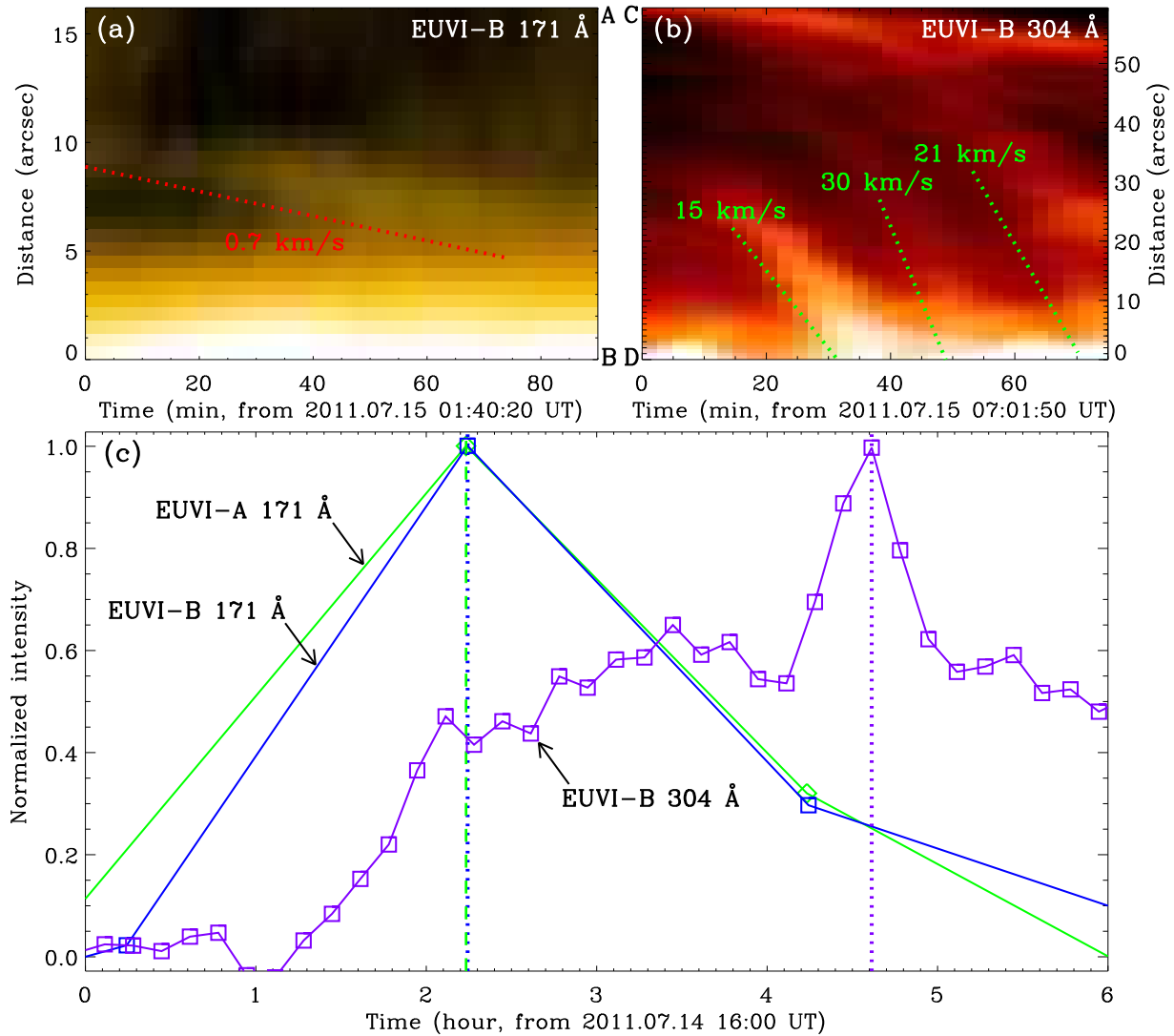}
\centering
\caption{Temporal evolution of the magnetic reconnection between open and closed structures and the coronal condensations observed by STEREO A and B. (a)-(b) Time slices of EUVI B 171 \AA~and 304 \AA~images along the AB and CD directions in the red rectangle and green box in Figures \ref{f:cc_stereo}(a) and (b), respectively. (c) Light curves of the EUVI B 171 \AA~(blue squares and line), B 304 \AA~(purple squares and line), and A 171 \AA~(green diamonds and line) channels in the purple and blue rectangles in Figures \ref{f:cc_stereo}(a)-(b) and (d), respectively. The red and green dotted lines in (a)-(b) separately show the motions of structures and condensations. The moving speeds are denoted by the numbers in (a)-(b). The vertical blue dotted, purple dotted, and green dashed lines in (c) mark the peaks of the EUVI B 171 \AA, B 304 \AA, and A 171 \AA~light curves, respectively; see the second condensation event in Table \ref{tab:measurements}. See Section \ref{sec:cc_stereo} for details.
\label{f:measurements_stereo}}
\end{figure}

\begin{figure}[ht!]
\includegraphics[width=0.9\textwidth]{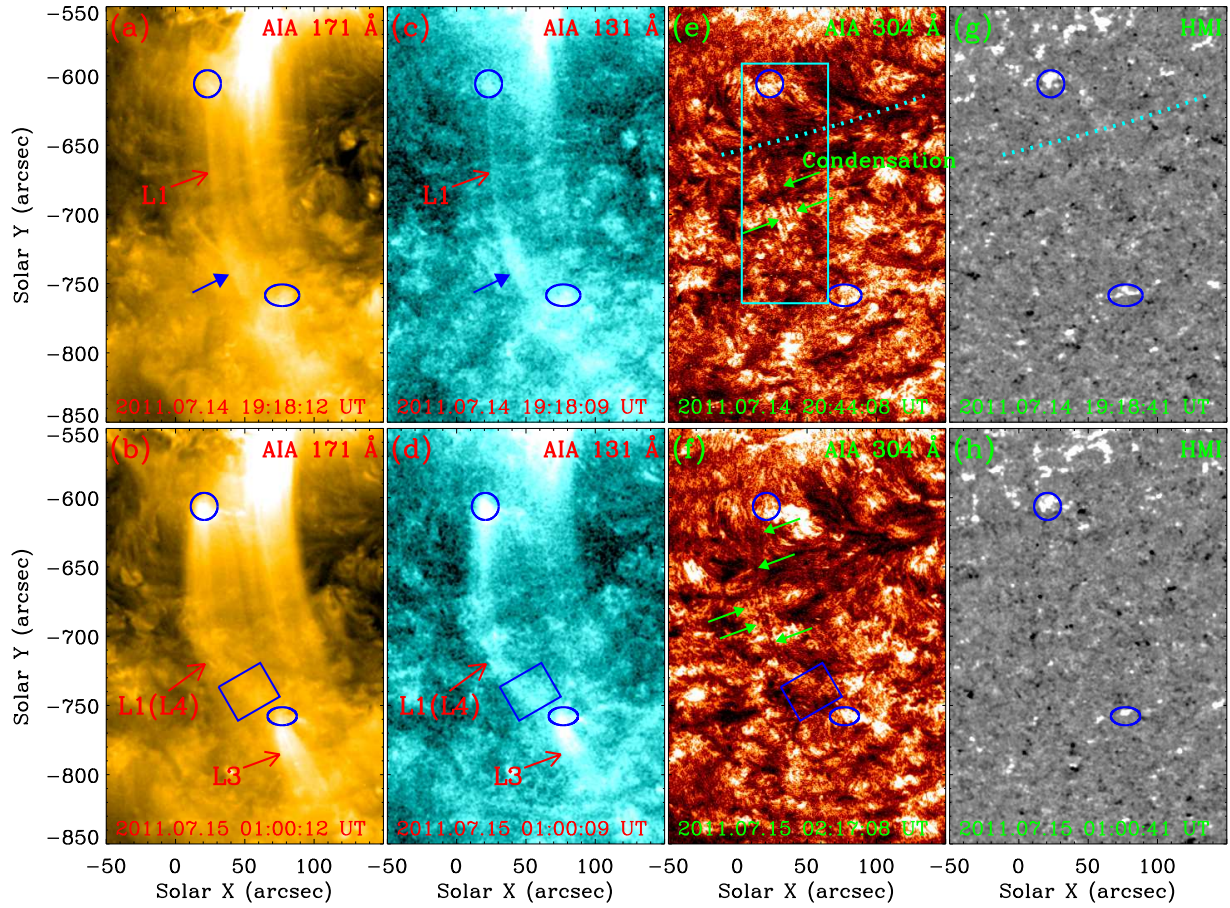}
\centering
\caption{Coronal condensations facilitated by magnetic reconnection between open and closed structures observed by SDO. (a)-(b) AIA 171 \AA, (c)-(d) 131 \AA, and (e)-(f) 304 \AA~images, and (g)-(h) HMI LOS magnetograms. Here, the AIA 304 \AA~images in (e)-(f) are enhanced employing the MGN technique. The blue circles and ellipses separately enclose the endpoints of structures L1 and L3. The blue solid arrows in (a) and (c) mark the bright emission in the AIA 171 \AA~and 131 \AA~channels, and the green solid arrows in (e)-(f) denote the condensations. The cyan dotted lines in (e) and (g) represent the filament underneath the structures L1. The cyan rectangle in (e) shows the FOV of Figures \ref{f:measurements_sdo}(a)-(b). The blue rectangles in (b), (d), and (f) mark the regions for the light curves of the AIA 171 \AA, 131 \AA, and 304 \AA~channels as displayed in Figure \ref{f:lightcurves_sdo} by the red, green, and blue lines. The FOV is denoted by the blue rectangle in Figure \ref{f:general}(c). An animation of the unannotated SDO observations is available. It covers 16 hr starting at 13:00 UT on 2011 July 14, and the time cadence is 1 minute. See Section \ref{sec:cc_sdo} for details. (An animation of this figure is available.)
\label{f:cc_sdo}}
\end{figure}

\begin{figure}[ht!]
\includegraphics[width=0.5\textwidth]{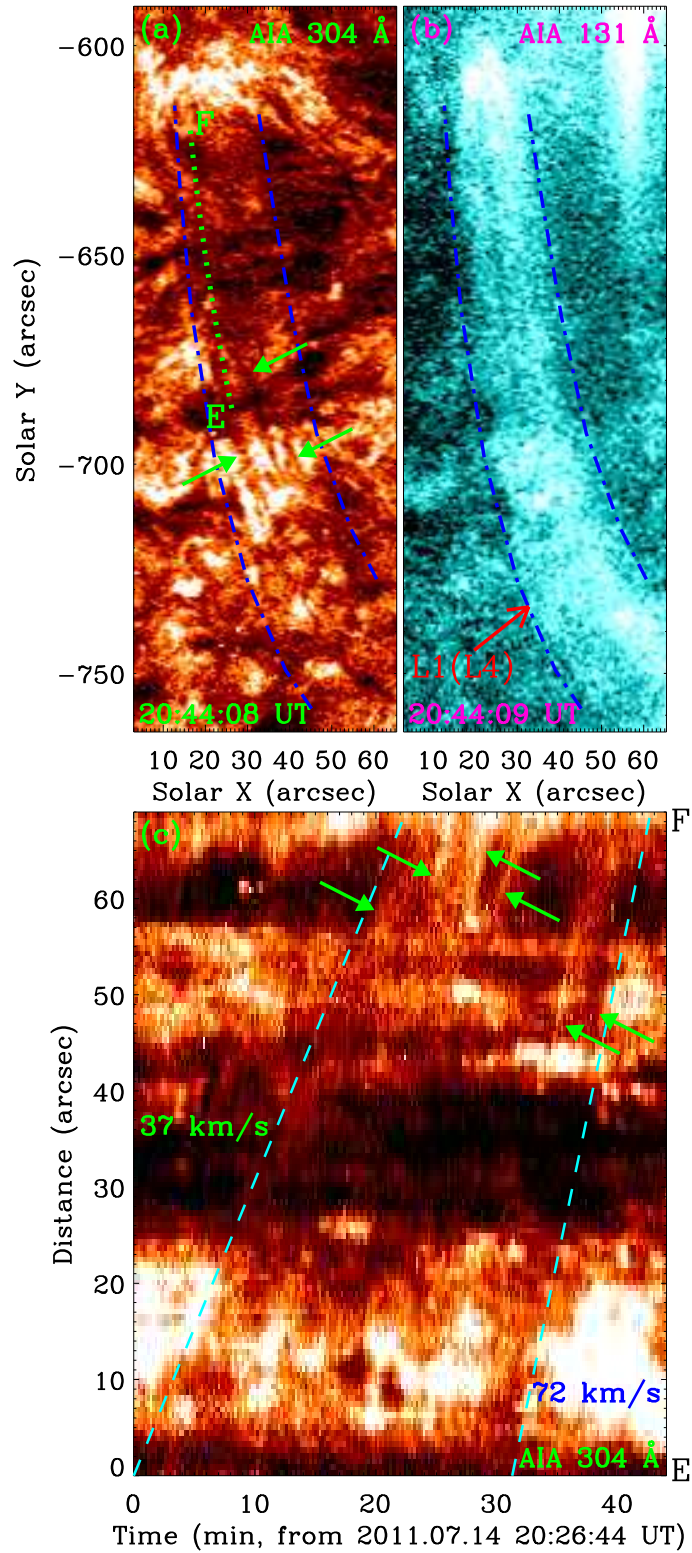}
\centering
\caption{Temporal evolution of the coronal condensations facilitated by magnetic reconnection between open and closed structures observed by SDO. (a) AIA 304 \AA~and (b) 131 \AA~images, and (c) a time slice of AIA 304 \AA~images along the green dotted line EF in (a). Here, the AIA 304 \AA~image in (a) is enhanced using the MGN technique. Same as in Figures \ref{f:cc_sdo}(e)-(f), the green solid arrows in (a) and (c) denote the condensations. The blue dash-dotted lines in (a)-(b) enclose the structures L1 and L4 in (b). The cyan dashed lines in (c) enclose the motions of condensations. The moving speeds are denoted by the numbers in (c). The FOV of (a)-(b) is denoted by the cyan rectangle in Figure \ref{f:cc_sdo}(e). An animation of the AIA 304 \AA~images (panel (a)) is available. It covers 80 minutes starting at 20:00 UT on 2011 July 14, and the time cadence is 1 minute. See Section \ref{sec:cc_sdo} for details. (An animation of this figure is available.)
\label{f:measurements_sdo}}
\end{figure}

Similar coronal condensation events facilitated by reconnection between open and closed structures above the limb observed by STEREO are presented (see Section \ref{sec:cc_stereo}). The repeated condensation events last for 1.5-10.2 hr, consistent with the lifetimes of condensations in \citet{2018ApJ...864L...4L, 2018ApJ...868L..33L, 2019ApJ...884...34L, 2020ApJ...L}. The moving speed (0.7 km s$^{-1}$) of structures L1 and the falling speed (15-30 km s$^{-1}$) of condensations are, however, smaller than those in \citet{2018ApJ...864L...4L, 2018ApJ...868L..33L, 2019ApJ...884...34L, 2020ApJ...L}. This may be caused by the longer time cadences of EUVI A and B 171 \AA~and 304 \AA~images than that of the AIA 171 \AA~and 304 \AA~images, see Section \ref{sec:obs}. Identical with \citet{2019ApJ...884...34L, 2020ApJ...L}, a filament is located under the higher-lying open structures L1, see Figures \ref{f:general} and \ref{f:cartoon}. This kind of magnetic system, consisting of the higher-lying open structures and lower-lying filament, may be prone to the condensations facilitated by reconnection between open and closed structures.

\begin{figure}[ht!]
\includegraphics[width=0.9\textwidth]{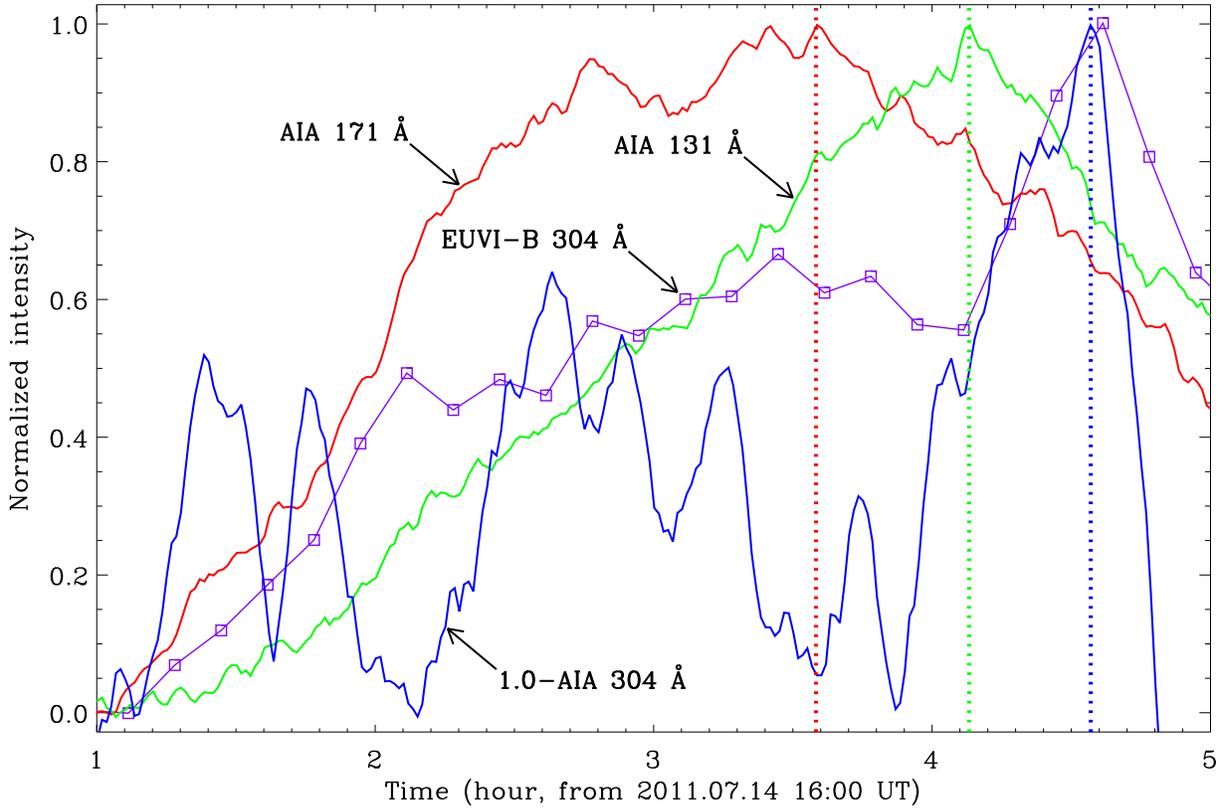}
\centering
\caption{The AIA 171 \AA~and 131 \AA~light curves (red and green lines) and the inverse of the AIA 304 \AA~light curve (blue line) in the blue rectangles in Figures \ref{f:cc_sdo}(b), (d), and (f). The red, green, and blue vertical dotted lines separately mark the peaks of the AIA 171 \AA, 131 \AA, and 304 \AA~light curves; see the second condensation event in Table \ref{tab:measurements}. Similar to Figure \ref{f:measurements_stereo}(c), the purple squares and line represent  the EUVI B 304 \AA~light curve. See Section \ref{sec:cc_sdo} for details.
\label{f:lightcurves_sdo}}
\end{figure}

\begin{figure}[ht!]
\includegraphics[width=0.96\textwidth]{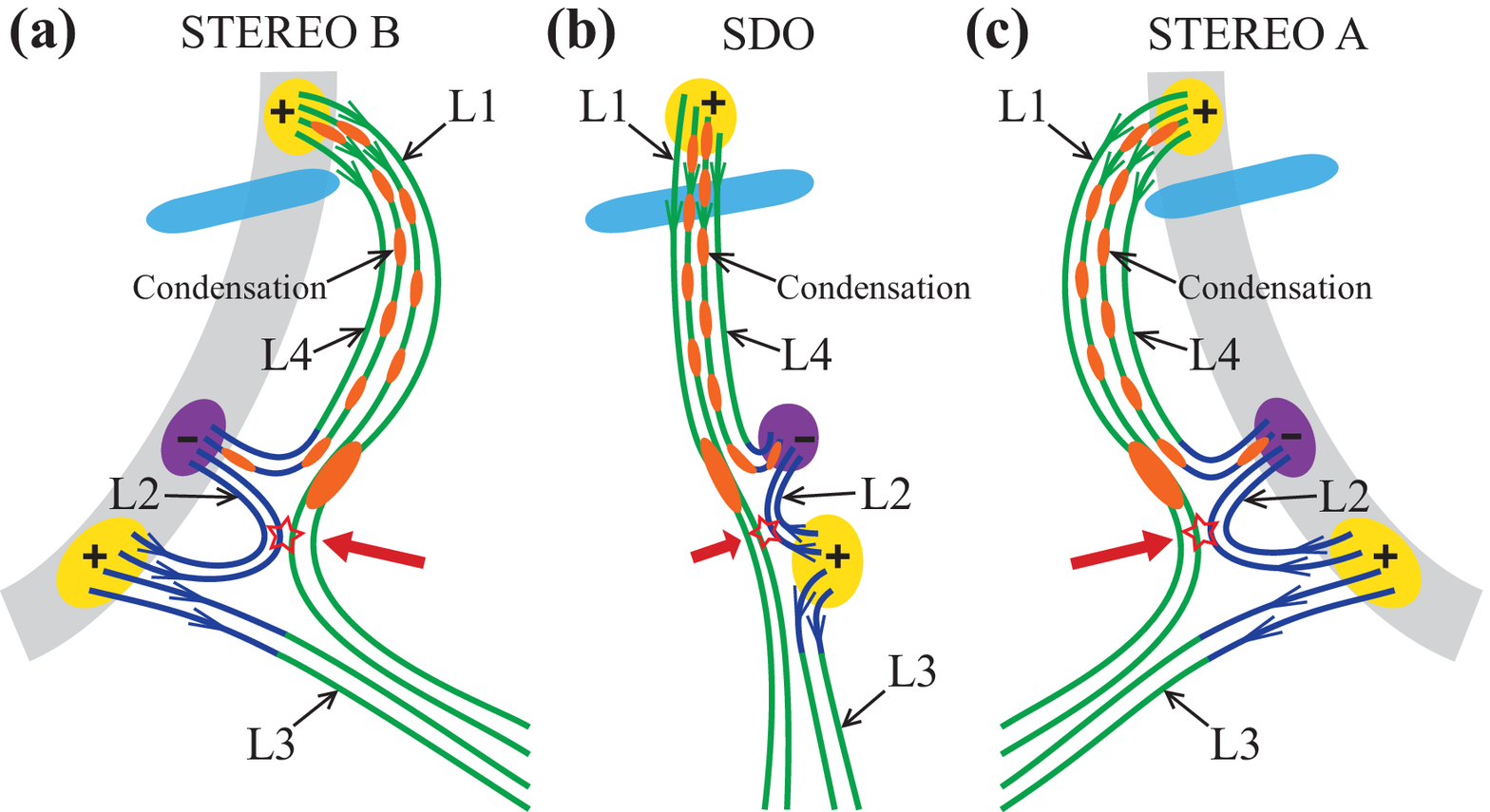}
\centering
\caption{Schematic diagrams of the coronal condensations facilitated by magnetic reconnection between open and closed structures as observed from three vantage points: (a) STEREO B, (b) SDO, and (c) STEREO A. The gray thick lines in (a) and (c) denote the solar limb. In (a)-(c), the yellow and purple ellipses with plus and minus signs represent the positive and negative magnetic fields, respectively. The green, blue, and green-blue lines separately show the magnetic field lines of structures L1, L2, L3, and L4, whose directions are marked by the green and blue arrows. The red solid arrows denote the moving directions of structures L1, and also mark the magnetic dips. The red stars represent the magnetic reconnection between structures L1 and L2. The  orange ellipses show the condensations. The cyan thick lines denote the filament underneath the higher-lying open structures L1. See Section \ref{sec:sum} for details.
\label{f:cartoon}}
\end{figure}

On-disk condensation events facilitated by reconnection between open and closed structures are first reported (see Section \ref{sec:cc_sdo}). Similar to those off-limb condensations facilitated by reconnection previously investigated \citep{2018ApJ...864L...4L, 2018ApJ...868L..33L, 2019ApJ...884...34L, 2020ApJ...L}, bright emission in AIA 171 \AA~and 131 \AA~images is detected on the disk during the cooling and condensation process. However, the AIA 171 \AA~structures are difficult to observe on the disk before the reconnection takes place. Moreover, dark, rather than bright, condensations are identified in AIA 304 \AA~images on the disk, consistent with the on-disk quiescent coronal rain in AR closed loops in H$\alpha$ against a bright background \citep{2012ApJ...745..152A, 2012SoPh..280..457A, 2014SoPh..289.4117A}. The moving speed (37-72 km s$^{-1}$) of on-disk condensations in AIA 304 \AA~images is larger than that (15-30 km s$^{-1}$) of off-limb condensations in EUVI 304 \AA~images, but identical to those of off-limb condensations in AIA 304 \AA~images \citep{2018ApJ...864L...4L, 2018ApJ...868L..33L, 2019ApJ...884...34L, 2020ApJ...L}, due to the time cadences of respective observations. Even though the structures and condensations on the disk are affected by the underneath background emission, similar cooling times from $\sim$0.9 MK to $\sim$0.6 MK (0.13-1.05 hr), and then to $\sim$0.05 MK (0.43-2.75 hr) are clearly identified to those of the off-limb bright condensations facilitated by reconnection in \citet{2018ApJ...864L...4L, 2018ApJ...868L..33L, 2019ApJ...884...34L, 2020ApJ...L} (see Figure \ref{f:lightcurves_sdo} and Table \ref{tab:measurements}).

On-disk condensations facilitated by reconnection between open and closed structures at high coronal altitudes are suggested to explain some of the flows, e.g., coronal rain, in chromospheric lines on the disk. Recently, employing IRIS and AIA observations, we proposed that some of the off-limb coronal rain events with loop-like paths in transition region and chromospheric lines originate from the condensations facilitated by interchange reconnection \citep{2020ApJ...L}. In this study, the coronal plasma would cool all the way down to the chromospheric temperatures, as reported in \citet{2020ApJ...L}, the chromospheric flows associated with the downflows of condensations then could be observed in the chromospheric lines. Using H$\alpha$ line center images of the full-disk H$\alpha$ telescope at the Huairou Solar Observing Station, chromospheric flows associated with the on-disk condensations are examined. On account of the width of condensations ($\sim$1.4\arcsec), the larger spatial sampling of H$\alpha$ images ($\sim$0.9\arcsec/pixel), and the influencing factors of ground-based telescopes, e.g., cloud and atmospheric turbulence, no associated chromospheric flow is evidently detected. To verify our suggestion, higher spatial resolution chromospheric observations, e.g., data from the New Vacuum Solar Telescope \citep[NVST;][]{2014RAA....14..705L} or the Daniel K. Inouye Solar Telescope (DKIST), are needed in the future works. 

Similar results of the on-disk condensations are obtained to those of the on-disk quiescent coronal rain in AR closed loops in H$\alpha$ (see Section \ref{sec:cc_sdo}). If we would consider the AIA 304 \AA~observations alone, the on-disk condensation, and hence coronal rain events we report here would resemble those occurring in magnetically closed field lines \citep{2012ApJ...745..152A, 2012SoPh..280..457A, 2014SoPh..289.4117A}, which are generally  interpreted as a manifestation of the heating-condensation cycles due to thermal non-equilibrium \citep{2001SoPh..198..325S, 2003A&A...411..605M, 2004A&A...424..289M, 2020PPCF...62a4016A}. However, combining the observations of SDO and STEREO A and B at different viewing angles, we find that the on-disk coronal rain in this study corresponds to the downflows of condensations facilitated by reconnection between open and closed structures. Similarly, if we analyse only the on-disk observations as recorded by the AIA as shown in Figure\,\ref{f:lightcurves_sdo}, it appears that the structures gradually brighten first in the 171\,\AA\ channel and then in the 131\,\AA\ channel. Such sequential appearance of coronal structures in the AIA channels sensitive to emission from progressively cooler plasma could be interpreted as a case of loop cooling after nanoflare heating \citep[e.g.,][]{2012ApJ...753...35V, 2015A&A...583A.109L}, for instance. In contrast, by employing observations from multiple vantage points, we show that the on-disk structure brightening in AIA images is due actually to cooling and condensation of plasma facilitated by reconnection in the high corona that is not necessarily regulated by any heating mechanism. To search for the origination of structures harbouring on-disk flow, e.g., coronal rain, in the transition region or chromospheric lines,  observations from different viewing angles are thus quite important. If there is no observation from other viewing angles, evolution of the associated structures, that may be difficult to observe on the disk, in multi-wavelength images during the cooling and condensation process needs to be examined, because the condensation facilitated by reconnection cools down from $\sim$0.9 MK, the characteristic temperature of the 171 \AA~channel, rather than from the higher temperatures.

\clearpage
\startlongtable
\begin{deluxetable}{c c c c c c c c c c c}
\tabletypesize{\scriptsize}
\tablecaption{General information on coronal condensations facilitated by magnetic reconnection between coronal open and closed structures observed by STEREO A and B from 2011 July 14 to 15. 
\label{tab:information}}
\tablehead{
 & \colhead{Start Time}  & \colhead{End Time} & \multicolumn{2}{c}{Appearance Time (UT)} & &  \multicolumn{2}{c}{Disappearance Time (UT)}  &  & \multicolumn{2}{c}{Lifetime (hr)}  \\
\cline{4-5}
\cline{7-8}
\cline{10-11}
Event & (UT) & (UT) & EUVI-A & EUVI-B & & EUVI-A & EUVI-B &  & EUVI-A & EUVI-B
}
\startdata
1 & 14 00:00 & 14 13:00 & 14 02:16 $\pm$0.2 hr  & 14 01:46 $\pm$0.2 hr &  & 14 11:56 $\pm$0.2 hr & 14 11:56 $\pm$0.2 hr & & 9.7$\pm$0.4 & 10.2$\pm$0.4 \\ 
2 & 14 13:00 & 14 21:26 & 14 19:56 $\pm$0.3 hr  & 14 19:56 $\pm$0.2 hr &  & 14 22:16 $\pm$2 hr & 14 21:26 $\pm$0.2 hr & & 2.3$\pm$2.3 & 1.5$\pm$0.4 \\ 
3 & 14 21:26 & 15 05:00 & 14 22:16 $\pm$2 hr  & 14 21:26 $\pm$0.2 hr &  & 15 04:16 $\pm$2 hr & 15 04:06 $\pm$0.1 hr & & 6$\pm$4 & 6.7$\pm$0.3 \\ 
4 & 15 05:00 & 15 12:00 & 15 08:16 $\pm$2 hr  & 15 06:39 &  & 15 12:16 $\pm$2 hr & 15 10:26 $\pm$0.2 hr & & 4$\pm$4 & 3.8$\pm$0.2 \\ 
5 & 15 12:00 & 15 23:59 & 15 14:48 $\pm$0.5 hr  & 15 14:36 $\pm$0.2 hr &  & 15 21:56 $\pm$0.3 hr & 15 21:56 $\pm$0.2 hr & & 7.1$\pm$0.8 & 7.3$\pm$0.4 \\
\enddata
\tablecomments{The 14 and 15 in the front of the columns 2-7 represent the dates of 2011 July 14 and 15, respectively. The appearance time, disappearance time, and lifetime of condensations are obtained from EUVI A and B 304 \AA~images.}
\end{deluxetable}

\startlongtable
\begin{deluxetable}{c c c c c c c c c }
\tabletypesize{\scriptsize}
\tablecaption{Temporal details of coronal condensations facilitated by reconnection between open and closed structures observed by STEREO A and B and SDO from 2011 July 14 to 15. 
\label{tab:measurements}}
\tablehead{
 &  & \multicolumn{3}{c}{Peak Time (UT)} & &  \multicolumn{3}{c}{Cooling Time (hr)}  \\
\cline{3-5}
\cline{7-9}
Event & Instrument & 171 \AA & 131 \AA & 304 \AA & & 171 \AA~to 131 \AA & 131 \AA~to 304 \AA & 171 \AA~to 304 \AA
}
\startdata
 & EUVI-A & - & -  & 14 04:26 $\pm$0.2 hr &  & - & - & -  \\ 
1   & EUVI-B & 14 02:14 $\pm$2 hr & -  & 14 04:26 $\pm$0.2 hr &  & - & - & 2.2$\pm$2.2  \\ 
  & AIA & 14 01:49 & 14 02:01  & 14 04:46 &  & 0.2 & 2.75 & 2.95 \\ 
\hline
 & EUVI-A & 14 18:14 $\pm$4.3 hr & -  & - &  & - & - & - \\ 
2  & EUVI-B & 14 18:14 $\pm$2 hr & -  & 14 20:36 $\pm$0.2 hr &  & - & - & 2.4$\pm$2.2 \\ 
  & AIA & 14 19:35 & 14 20:08  & 14 20:34 &  & 0.55 & 0.43 & 0.98  \\
\hline
 & EUVI-A & 15 00:14 $\pm$2 hr & -  & - &  & - & - & - \\ 
3  & EUVI-B & 15 00:14 $\pm$0.7 hr & -  & 15 01:51 &  & - & - & 1.6$\pm$0.7 \\ 
  & AIA & 15 00:23 & 15 00:33  & 15 01:59 &  & 0.17 & 1.43 & 1.6  \\
\hline
 & EUVI-A & - & -  & 15 08:16 $\pm$2 hr &  & - & - & - \\ 
4  & EUVI-B & - & -  & 15 07:54 &  & - & - & - \\ 
  & AIA & 15 06:25 & 15 06:33  & 15 07:56 &  & 0.13 & 1.38 & 1.52  \\
\hline
 & EUVI-A & - & -  & 15 16:46 $\pm$0.2 hr &  & - & - & - \\ 
5  & EUVI-B & 15 14:14 $\pm$2 hr & -  & 15 16:26 $\pm$0.2 hr &  & - & - & 2.2$\pm$2.2 \\ 
  & AIA & 15 13:52 & 15 14:55  & 15 16:09 &  & 1.05 & 1.23 & 2.28  \\         
\enddata
\tablecomments{Similar to Table \ref{tab:information}, the 14 and 15 in the front of the columns 3-5 separately represent the dates of 2011 July 14 and 15. For the second event, the peak times of the STEREO/EUVI A 171 \AA, B 171 \AA, and B 304 \AA~light curves are marked by the vertical green dashed, blue dotted, and purple dotted lines in Figure \ref{f:measurements_stereo}(c), and those of the SDO/AIA 171 \AA, 131 \AA, and 304 \AA~light curves are denoted by the red, green, and blue vertical dotted lines in Figure \ref{f:lightcurves_sdo}.}
\end{deluxetable}

\acknowledgments
The authors thank the anonymous referee for helpful comments. We are indebted to the STEREO and SDO teams for providing the data. AIA data are the courtesy of NASA/SDO and the AIA, EVE, and HMI science teams. This work is supported by the Strategic Priority Research Program of Chinese Academy of Sciences, Grant No. XDB 41000000, the National Natural Science Foundations of China (12073042, 11873059, and 11773039), and the Key Research Program of Frontier Sciences (ZDBS-LY-SLH013) and the Key Programs (QYZDJ-SSW-SLH050) of Chinese Academy of Sciences. H. Q. S. is supported by the National Natural Science Foundation of China (U2031109). We acknowledge the usage of JHelioviewer software \cite[][]{2017A&A...606A..10M}. This research has made use of NASA's Astrophysics Data System. 




\begin{thebibliography}{}

\bibitem[Ahn et al.(2014)]{2014SoPh..289.4117A} Ahn, K., Chae, J., Cho, K.-S., et al.\ 2014, \solphys, 289, 4117
\bibitem[Antolin(2020)]{2020PPCF...62a4016A} Antolin, P.\ 2020, Plasma Physics and Controlled Fusion, 62, 014016
\bibitem[Antolin \& Rouppe van der Voort(2012)]{2012ApJ...745..152A} Antolin, P., \& Rouppe van der Voort, L.\ 2012, \apj, 745, 152
\bibitem[Antolin et al.(2012)]{2012SoPh..280..457A} Antolin, P., Vissers, G., \& Rouppe van der Voort, L.\ 2012, \solphys, 280, 457
\bibitem[Auch{\`e}re et al.(2018)]{2018ApJ...853..176A} Auch{\`e}re, F., Froment, C., Soubri{\'e}, E., et al.\ 2018, \apj, 853, 176
\bibitem[Berger et al.(2012)]{2012ApJ...758L..37B} Berger, T.~E., Liu, W., \& Low, B.~C.\ 2012, \apjl, 758, L37
\bibitem[Cheng et al.(2018)]{2018ApJ...866...64C} Cheng, X., Li, Y., Wan, L.~F., et al.\ 2018, \apj, 866, 64
\bibitem[Chitta et al.(2018)]{2018A&A...615L...9C} Chitta, L.~P., Peter, H., \& Solanki, S.~K.\ 2018, \aap, 615, L9
\bibitem[De Pontieu et al.(2014)]{2014SoPh..289.2733D} De Pontieu, B., Title, A.~M., Lemen, J.~R., et al.\ 2014, \solphys, 289, 2733
\bibitem[Fang et al.(2013)]{2013ApJ...771L..29F} Fang, X., Xia, C., \& Keppens, R.\ 2013, \apjl, 771, L29
\bibitem[Field(1965)]{1965ApJ...142..531F} Field, G.~B.\ 1965, \apj, 142, 531
\bibitem[Froment et al.(2020)]{2020A&A...633A..11F} Froment, C., Antolin, P., Henriques, V.~M.~J., et al.\ 2020, \aap, 633, A11
\bibitem[Froment et al.(2017)]{2017ApJ...835..272F} Froment, C., Auch{\`e}re, F., Aulanier, G., et al.\ 2017, \apj, 835, 272
\bibitem[Froment et al.(2018)]{2018ApJ...855...52F} Froment, C., Auch{\`e}re, F., Miki{\'c}, Z., et al.\ 2018, \apj, 855, 52
\bibitem[Howard et al.(2008)]{2008SSRv..136...67H} Howard, R.~A., Moses, J.~D., Vourlidas, A., et al.\ 2008, \ssr, 136, 67
\bibitem[Kaiser et al.(2008)]{2008SSRv..136....5K} Kaiser, M.~L., Kucera, T.~A., Davila, J.~M., et al.\ 2008, \ssr, 136, 5
\bibitem[Kaneko \& Yokoyama(2015)]{2015ApJ...806..115K} Kaneko, T., \& Yokoyama, T.\ 2015, \apj, 806, 115
\bibitem[Kaneko \& Yokoyama(2017)]{2017ApJ...845...12K} Kaneko, T., \& Yokoyama, T.\ 2017, \apj, 845, 12
\bibitem[Kohutova et al.(2020)]{2020A&A...639A..20K} Kohutova, P., Antolin, P., Popovas, A., et al.\ 2020, \aap, 639, A20
\bibitem[Kohutova et al.(2019)]{2019A&A...630A.123K} Kohutova, P., Verwichte, E., \& Froment, C.\ 2019, \aap, 630, A123
\bibitem[Lemen et al.(2012)]{2012SoPh..275...17L} Lemen, J.~R., Title, A.~M., Akin, D.~J., et al.\ 2012, \solphys, 275, 17
\bibitem[Li et al.(2016a)]{2016Ap&SS.361..301L} Li, D., Ning, Z., \& Su, Y.\ 2016a, \apss, 361, 301
\bibitem[Li et al.(2015)]{2015A&A...583A.109L} Li, L.~P., Peter, H., Chen, F., et al.\ 2015, \aap, 583, A109
\bibitem[Li et al.(2019)]{2019ApJ...884...34L} Li, L. P., Peter, H., Chitta, L. P., et al.\ 2019, \apj, 884, 34
\bibitem[Li et al.(2020)]{2020ApJ...L} Li, L. P., Peter, H., Chitta, L. P., \& Song, H. Q.\ 2020, \apj, 905, 26
\bibitem[Li \& Zhang(2009)]{2009ApJ...703..877L} Li, L. P., \& Zhang, J.\ 2009, \apj, 703, 877
\bibitem[Li et al.(2016b)]{2016NatPh..12..847L} Li, L. P., Zhang, J., Peter, H., et al.\ 2016b, Nature Physics, 12, 847
\bibitem[Li et al.(2018a)]{2018ApJ...864L...4L} Li, L. P., Zhang, J., Peter, H., et al.\ 2018a, \apjl, 864, L4
\bibitem[Li et al.(2018b)]{2018ApJ...868L..33L} Li, L. P., Zhang, J., Peter, H., et al.\ 2018b, \apjl, 868, L33
\bibitem[Li et al.(2016c)]{2016ApJ...829L..33L} Li, L.~P., Zhang, J., Su, J.~T., et al.\ 2016c, \apjl, 829, L33
\bibitem[Liu et al.(2010)]{2010ApJ...723L..28L} Liu, R., Lee, J., Wang, T., et al.\ 2010, \apjl, 723, L28
\bibitem[Liu et al.(2012)]{2012ApJ...745L..21L} Liu, W., Berger, T.~E., \& Low, B.~C.\ 2012, \apjl, 745, L21
\bibitem[Liu et al.(2014)]{2014RAA....14..705L} Liu, Z., Xu, J., Gu, B.-Z., et al.\ 2014, Research in Astronomy and Astrophysics, 14, 705-718
\bibitem[Mackay et al.(2010)]{2010SSRv..151..333M} Mackay, D.~H., Karpen, J.~T., Ballester, J.~L., et al.\ 2010, \ssr, 151, 333
\bibitem[Mason et al.(2019)]{2019ApJ...874L..33M} Mason, E.~I., Antiochos, S.~K., \& Viall, N.~M.\ 2019, \apjl, 874, L33
\bibitem[Morgan \& Druckm{\"u}ller(2014)]{2014SoPh..289.2945M} Morgan, H., \& Druckm{\"u}ller, M.\ 2014, \solphys, 289, 2945
\bibitem[M{\"u}ller et al.(2003)]{2003A&A...411..605M} M{\"u}ller, D.~A.~N., Hansteen, V.~H., \& Peter, H.\ 2003, \aap, 411, 605
\bibitem[{{M{\"u}ller} {et~al.}(2017){M{\"u}ller}, {Nicula}, {Felix},
  {Verstringe}, {Bourgoignie}, {Csillaghy}, {Berghmans}, {Jiggens},
  {Garc{\'{\i}}a-Ortiz}, {Ireland}, {Zahniy}, \& {Fleck}}]{2017A&A...606A..10M}
{M{\"u}ller}, D. A. N., {Nicula}, B., {Felix}, S., {et~al.} 2017, \aap, 606, A10
\bibitem[M{\"u}ller et al.(2004)]{2004A&A...424..289M} M{\"u}ller, D.~A.~N., Peter, H., \& Hansteen, V.~H.\ 2004, \aap, 424, 289
\bibitem[Oliver et al.(2016)]{2016ApJ...818..128O} Oliver, R., Soler, R., Terradas, J., et al.\ 2016, \apj, 818, 128
\bibitem[Parker(1953)]{1953ApJ...117..431P} Parker, E.~N.\ 1953, \apj, 117, 431
\bibitem[Pesnell et al.(2012)]{2012SoPh..275....3P} Pesnell, W.~D., Thompson, B.~J., \& Chamberlin, P.~C.\ 2012, \solphys, 275, 3
\bibitem[Pneuman(1983)]{1983SoPh...88..219P} Pneuman, G.~W.\ 1983, \solphys, 88, 219
\bibitem[Priest \& Forbes(2000)]{2000mrmt.conf.....P} Priest, E., \& Forbes, T.\ 2000, Magnetic Reconnection: MHD Theory and Applications / Eric Priest
\bibitem[Reeves et al.(2015)]{2015ApJ...807....7R} Reeves, K.~K., McCauley, P.~I., \& Tian, H.\ 2015, \apj, 807, 7
\bibitem[Schou et al.(2012)]{2012SoPh..275..229S} Schou, J., Scherrer, P.~H., Bush, R.~I., et al.\ 2012, \solphys, 275, 229
\bibitem[Schrijver(2001)]{2001SoPh..198..325S} Schrijver, C.~J.\ 2001, \solphys, 198, 325
\bibitem[Shen et al.(2019)]{2019ApJ...883..104S} Shen, Y., Qu, Z., Yuan, D., et al.\ 2019, \apj, 883, 104
\bibitem[Tian et al.(2014)]{2014ApJ...797L..14T} Tian, H., Li, G., Reeves, K.~K., et al.\ 2014, \apjl, 797, L14
\bibitem[Vashalomidze et al.(2015)]{2015A&A...577A.136V} Vashalomidze, Z., Kukhianidze, V., Zaqarashvili, T.~V., et al.\ 2015, \aap, 577, A136
\bibitem[Viall \& Klimchuk(2012)]{2012ApJ...753...35V} Viall, N.~M. \& Klimchuk, J.~A.\ 2012, \apj, 753, 35
\bibitem[Xia et al.(2014)]{2014ApJ...792L..38X} Xia, C., Keppens, R., Antolin, P., et al.\ 2014, \apjl, 792, L38
\bibitem[Yan et al.(2018)]{2018ApJ...853L..18Y} Yan, X.~L., Yang, L.~H., Xue, Z.~K., et al.\ 2018, \apjl, 853, L18

\end{thebibliography}
\end{document}